\documentclass[onecolumn]{aa}  
\usepackage{graphicx, natbib} 
\usepackage{txfonts}
\def \msun{{\rm ~M_{\odot}}}
\def \mdot{\dot {\rm M}}

\newcommand{\swift}{{\it SWIFT}}
\newcommand{\sax}{{\it BeppoSAX}}
\newcommand{\asca}{{\it ASCA}}
\newcommand{\xmm}{{\it XMM-Newton}}

\newcommand{\integral}{{\it INTEGRAL}}
\newcommand{\chandra}{{\it Chandra}}
\newcommand{\mc}{\multicolumn}
\begin{document}
\titlerunning{The broad band X-ray spectra of the UCXBs}
   \title{The {\it INTEGRAL}{\thanks
{{\it INTEGRAL\/} is an ESA project with instruments and Science Data Center funded by ESA member states (especially the PI countries: Denmark, France, Germany, Italy, Switzerland, Spain), Czech Republic and Poland, and with the participation of Russia and the USA.
}} long monitoring of persistent Ultra Compact X-ray Bursters 
}
   \subtitle{}

   \author{M. Fiocchi\inst{1}, A. Bazzano\inst{1}, P. Ubertini 
          \inst{1},
	A. J. Bird \inst{2},
          L. Natalucci \inst{1},
V. Sguera \inst{3}	
          }

   \offprints{}

   \institute{Instituto di Astrofisica Spaziale e Fisica Cosmica di Roma (INAF). Via Fosso del Cavaliere 100, Roma, I-00133, Italy\\
\email{mariateresa.fiocchi@iasf-roma.inaf.it}
 \and {School of Physics and Astronomy, University of Southampton, SO17 1BJ, UK}
 \and { Instituto di Astrofisica Spaziale e Fisica Cosmica di Bologna (INAF). Via Gobetti 101, Bologna, I-40129, Italy}  
            }

   \date{Received ..........., accepted ............}

% \abstract{}{}{}{}{} 
% 5 {} token are mandatory
 
  \abstract
   {The combination of compact objects, short period variability and 
peculiar chemical composition of the Ultra Compact X-ray Binaries make up a very
interesting laboratory to study accretion processes and 
thermonuclear burning on the neutron star surface.\\ 
The improved large optical telescopes and more sensitive X-ray satellites 
have increased the number of known Ultra Compact X-ray Binaries allowing their 
study with unprecedented detail.
   }
   {We analyze the average properties common to 
all ultra compact Bursters observed by \integral\/ 
from $\sim0.2$~keV to $\sim150$~keV.  
   }
   {We have performed a systematic analysis of the \integral\/ public data 
and Key-Program proprietary observations of
a sample of the Ultra Compact X-ray Binaries. In order to study their average properties in a very
broad energy band, we combined \integral\/ with \sax\/ and \swift\/ data whenever possible.
For sources not showing any significant flux variations along the \integral\/ monitoring,
 we build 
the average spectrum by combining all available data; in the case of variable fluxes, we 
use simultaneous \integral\/ and \swift\/ observations when available. Otherwise we compared 
IBIS and PDS data to check the variability and combine \sax\/ with \integral\//IBIS data.
   }
   {
All spectra are well represented by a two component model consisting
of a disk-blackbody and Comptonised emission.
The majority of these compact sources spend most of the time in a canonical low/hard state,
with dominating Comptonised component and
accretion rate $\mdot$ lower than $\sim 10^{-9}\msun/yr$, not depending on the used model to fit the data.\\}


   \maketitle
%
%________________________________________________________________

\section{Introduction}
Ultracompact X-ray binaries (UCXBs) are binaries with orbital periods ($P_{\rm orb}$)
shorter than $\approx$1~hr in which a neutron star or black
hole accrete matter from a companion low mass star.\\
 Their short periods
rule out ordinary hydrogen-rich companion stars, since these stars are
too big and do not fit in the Roche lobe (Nelson
et al. 1986).
UCXBs are rare objects and their identification is very difficult
because it is hard to measure $P_{\rm orb}$ 
in LMXBs.
Eight out of 52 LMXBs with measured orbital period are
in the ultracompact regime
(in 't Zand et al. 2007, Nelemans \& Jonker 2006). 
Another thirteen UCXBs have been identified by indirect methods 
depending on the notion that in an UCXB the
accretion disk must be relatively small. The first indirect method
concerns the ratio of optical to X-ray flux. 
\citeauthor{vm94} (1994),
assuming that most of the optical flux is generated by the X-ray irradiation of the
accretion disc,
showed that the absolute magnitudes of LMXBs correlate with the 
X-ray luminosities and the size of the disc.
Seven UCXBs 
have been identified following this method
(e.g., Juett et al. 2001). The second indirect method concerns the
critical accretion rate below which a system becomes transient 
(in t Zand et al. 2007); it allowed the identification of six new UCXB candidates.
\section{Sample}
In order to obtain a broad band energy spectra, we carried out a systematic analysis of all Bursters UCXBs
 reported in the latest 
\integral\//IBIS survey (Bird et al. 2007) having the following properties:
a) they are persistent UCXB Bursters; 
b) there are available data in the soft X-ray energy band (at least $\sim 2-8 keV$) from \swift\//XRT 
or \sax\/ archives;
c) when a source does not vay during \integral\/ monitoring,
we use all available X-ray data, while when it varies we make use 
only of IBIS and XRT simultaneous spectra.  
If simultaneous data are not available, PDS and IBIS data are used to verify possible variability.\\

\begin{table*}[t] 
\caption[]{Sample of UCXBs. The galactic $N_H$ value was estimated from the 21 cm measurements of Dickey \& Lockman, 1990}
\label{tab1}
\begin{flushleft}
%\scriptsize
\begin{tabular}{lcccccccc}
\hline
\multicolumn{9}{c}{\it \bf{Persistent Bursters}} \\
&&&&&&&&\\
Name                       & (1)   & Distance   & $P_{\rm orb}$ &N$_{H}$&$M_V$&donor&long burst&REF\\
                           &       & kpc        & minutes &$10^{21}~cm^{-2}$&&&&\\
&&&&&&&&\\
%Instrument & IBIS$_{exp}$ &Range\\
\hline
&&&&&&&&\\
\multicolumn{9}{c}{\it UCXBs with orbital periods} \\
&&&&&&&&\\
4U 1820-303 (in NGC 6624)  &  pX        & 5.8-6.4-7.6   & 11       &1.5&3.7&He dwarf 0.06-0.08M$\sun$&yes&a,q,c,j\\
4U 1850-087 (in NGC 6712)  &  pO        & 6.8-8.0       & 21 or 13 &2.8&5.1&dwarf 0.04M$\sun$& ...&b,r,f,k \\
%4U 1915-05                &  px        & 8.4-10.8$^{\rm c}$      & 50$^{\rm s}$       &2.6&&$>$5.4&&\\
&&&&&&&&\\
\hline
&&&&&&&&\\
\multicolumn{9}{c}{\it UCXBs with tentative orbital periods} \\
&&&&&&&&\\
4U 0614+091                &  pO,r      & 1.5-3.0      & 50 &5.5&3.7&C/O dwarf&yes &d,t,m,f,w\\
XB 1832-330  (in NGC 6652) &  pO        & 9.2-14.3      & 55 &1.2&$>$3.5&blue star $M_V$=3.7 or dwarf&...&e,u,p\\
&&&&&&&&\\
\hline
&&&&&&&&\\
\multicolumn{9}{c}{\it candidate UCXBs with low optical to X-ray flux} \\
&&&&&&&&\\
%2S 0918-549               &  r    &  4.2$^{\rm f}$            &&7.3&&\\
1A 1246-588                &  r    &  5              &...&4.0&$>$3.5& blue star $M_V>$3.5 &yes &g,r\\
4U 1812-12                 &  r    &  4.1           &...&7.3&...&weak blue star &...&m,h,g,x  \\
&&&&&&&&\\
\hline
&&&&&&&&\\
\multicolumn{9}{c}{\it candidate UCXBs based on method from in 't Zand et al. (2007)} \\
&&&&&&&&\\
SAX J1712.6-3739           &  z        &   5.9-7.9 &   ...    &13.4 &...&...&...&i,y\\
4U 1705-32                 &  z        &   11-15    &...      &2.9 &...&...&yes &l\\
%4U 1724-307 (in Terzan 2) & z         &   6.6-10$^{\rm m}$    &  ...   &6.6 &...&...&...\\
1RXS J172525.5-325717      & z         &   $<$14.5    & ...   &8.7 &...&...&yes &n,s\\
SLX 1735-269               & z         &   6-8.5       &...   &4.8 &...&...&yes &o,z\\
SLX 1737-282               & z         &   6.5          &...  &7.4 &...&...&yes &o,v,aa\\
%SLX 1744-299              & z         &   8.5 $^{\rm p}$          &... &...&... \\
&&&&&&&&\\
\hline
&&&&&&&&\\
\end{tabular}
\noindent\\
(1) Ultracompact identification based on:
    r = $L_{\rm x}/L_{\rm opt}$,
    p = period measurement (pX using X-ray data, pO using optical modulation), z=in't Zand method\\
%    x = persistent burster with low L; z=in't Zand method\\
$^{\rm a}$kuulkers et al. 2003, Shaposhnikov and Titarchuk 2004, Vacca et al. 1986;
$^{\rm b}$Harris et al. 1996, Paltinieri 2001;
$^{\rm c}$ Strohmayer \& Brown 2002;
%$^{\rm c}$Smale et al. 1998, Narita et al. 2003;
%$^{\rm c}$ Rappaport et al. 1987;
$^{\rm d}$Brandt et al. 1992;
$^{\rm e}$in 't Zand et al. 1998, Chaboyer et al. 2000;
%$^{\rm f}$Chenevez et al. 2007;
$^{\rm f}$ Homer et al. 1996;
$^{\rm g}$Bassa et al. 2006;
$^{\rm h}$Cocchi et al. 2000;
$^{\rm i}$Cocchi et al. 2001, Jonker et al. 2004;
$^{\rm j}$Bloser et al. 2000;
$^{\rm k}$Sidoli et al. 2006;
$^{\rm l}$in 't Zand et al. 2005;
%$^{\rm m}$Gladstone et al. 2006;
$^{\rm m}$Nelemans, Jonker et al. 2004;
$^{\rm n}$Cornelisse et al. 2002;
$^{\rm o}$Molkov et al. 2005;
%$^{\rm p}$Goldwurml et al. 2004;
$^{\rm p}$ Parmar et al. 2001;
$^{\rm q}$Stella et al. 1987;
%$^{\rm r}$Homer et al. 1996;
%$^{\rm s}$White \& Swank 1982;
%$^{\rm s}$ Bassa et al. 2006;
$^{\rm r}$ Levine et al 2006, Piro et al. 1997, Boller et al. 1997;
$^{\rm s}$ Chenevez et al. 2007;
$^{\rm t}$O'Brien et al. 2005;
$^{\rm u}$Heinke et al. 2001;
$^{\rm v}$ Falanga et al. 2007;
$^{\rm w}$Piraino et al. 1999;
$^{\rm x}$Barret et al. 2003;
$^{\rm y}$Fiocchi et al. 2008;
$^{\rm z}$Gladstone et al. 2007;
$^{\rm aa}$in't Zand et al. 2002;
%$^{\rm dd}$ in't Zand et al 2005;
%$^{\rm ff}$ Molkov et al. 2005;
\label{log}
\end{flushleft}
\end{table*}
Following these criteria, we collected high quality IBIS
average spectra with a long exposure time and 
build up a wide band spectrum from $\sim$0.5 to $\sim$150 keV.
As for the source SLX1735-269 there is evidence of variability in the IBIS data set
and we have instead used the archival simultaneous broad band \sax\/ data to derive spectral parameters.
An overview of the observational properties of the known and candidate
UCXBs is given in Table~\ref{tab1}, which 
lists their name, ultracompact identification method, distance,
orbital
period, galactic column density, 
absolute V-band
magnitude,
available information on the donor and finally the eventual detection of long intermediate burst or
superburst.
This sample represents $\sim 60\%$ of all known persistent burster ultra compact sources
and candidates.
\section{Observations and Data Analysis}
The \integral\/ (Winkler et al. 2003) 
data set consist of the IBIS/ISGRI (Ubertini et al. 2003) observations
from revolution 37 to 488. Images for each available pointing were generated 
using the scientific analysis software (OSA v6.0) 
released by the \integral\/ Scientific Data Center.
Count rates at the position of the sources were extracted from individual images in order to 
provide light curves; average fluxes were then extracted from such light curves and combined to produce
sources spectra (see Bird et al. 2007 for details).

\sax\/ LECS, MECS and PDS event files and spectra,
were generated by means of the Supervised Standard Science Analysis
\cite{fio99}.
Both LECS and MECS spectra were accumulated in circular regions
of 8' radius.
The PDS spectra were obtained
with the background rejection method based on fixed rise time thresholds.

XRT/\swift\/ data reduction was performed using Swift software version 2.7.1 released as part of the HEASOFT software.
Data were collected
in the Photon Counting mode (PC)
or in the Windowed Timing (WT) mode with the grade filtering 0-12 or 0-2, respectively.
For the PC mode, events extracted within a circular region 
of radius 20 pixels,
centered on the source position. For WT mode, events for spectral analysis were extracted within a box region
centered on the source position and with a width of 40 pixels and an height large enough to include
all the photons in the Y dimension. The background was selected using a region on either side 
of the source.

Table 2, available at the CDS, reports on the log of the observations.

The whole
data set was fitted with this physical model,
using XSPEC v.\ 11.3.1.
Errors are quoted at 90\% confidence level for one parameter of interest.
In the fitting procedure, a multiplicative constant has been introduced to take into account
the cross calibration mismatches between the soft X-ray and the \integral\/ data;
this constant has been found to be $\sim$1 for all sources, but for 1A1246-588 (see below).
% Table 2 available electronically only

\begin{table*}
\caption[]{Data observation details}
\label{obs}
\begin{flushleft}
\scriptsize
\begin{tabular}{lcccccccc}
\hline\hline
&&&&&&&&\\
Name & Expos.   &START & Expos. & START& Expos. &Expos. &Expos. & Energy range\\
%&&&&&&&&\\
& INTEGRAL/IBIS   &SWIFT/XRT & SWIFT/XRT & BeppoSAX& BeppoSAX/LECS &BeppoSAX/MECS &BeppoSAX/PDS & \\
&&&&&&&&\\
&ks&&ks&&ks&ks&ks&keV\\
&&&&&&&&\\
\hline
&&&&&&&&\\
4U 1820-303     &8&2007-03-16 &4&...&...&...&...&2.5-60 \\

                &...&...&...&1999-08-31 &22&46&20&0.5-50 \\
&&&&&&&&\\
4U 1850-087     &2060&...&...&1997-04-24 &18&42&18&0.5-150 \\
&&&&&&&&\\
4U 0614+091      &23 &...&...&1996-10-13&12 &45 &21 &0.5-150 \\
&&&&&&&&\\
XB 1832-330      &3067&...&...&2001-03-17&29&54&29&0.2-150 \\
&&&&&&&&\\
1A 1246-588      &1606&2006-08-11&12&...&...&...&...&0.3-150\\
&&&&&&&&\\
4U 1812-12       &2054&...&...&2000-04-20&15&33&14&0.3-200 \\
&&&&&&&&\\
SAX J1712.6-3739 &685&...&...&1999-09-01&11&21&10&0.5-150 \\
&&&&&&&&\\
4U 1705-32       &4832 &2007-02-28 &7 &...&...&...&...&0.3-100 \\
&&&&&&&&\\
1RXS J172525.5-32 &5309&2007-03-17&2.5&...&...&...&...&0.6-100\\
&&&&&&&&\\
SLX 1735-269     &...&...&...&1997-09-18&9&29&14&0.5-80 \\

                 &...&...&...&1999-08-29&15&37&17&0.3-80 \\
&&&&&&&&\\
SLX 1737-282     &6206&...&...&2000-10-14&14&48&...&0.5-70\\
&&&&&&&&\\
\hline\hline
\end{tabular}
\end{flushleft}
\end{table*}

\section{Spectral Analysis}
To analyze all data in a homogeneous way, 
we searched a model at work for the whole sample.
Firstly, average spectra have been fitted with a simple power law model, 
resulting in a poor fit to the
data with reduced chi squares $>$10 for all spectra.  
Generally, the strongest obtained residuals were at energy below few keV,
indicating that a more complex model with a soft component
was needed to fit these data.

For the soft component we used a multicolor disk blackbody in the formulation of
Makishima et al. \cite{m:86}.
The two parameters of this model are free:
${\rm r_{in}({\cos}~i)^{0.5}}$ where ${\rm r_{in}}$ is the innermost
radius of the disk, {\em i} is the inclination angle of the disk and
${\rm kT_{in}}$ the blackbody effective temperature at ${\rm r_{in}}$.\\
It is expected, however, that the true spectrum emitted by the
accretion disk will be that of a modified blackbody as a result of the
effects of electron scattering, and the multi color disk parameters must
then be modified by a spectral hardening factor $f$ (Ebisawa et al. 1994).
The effective temperature and inner radius are $kT_{eff}=kT_{in}/f$ and
$R_{eff}=f^{2}R_{in}$. Shimura \& Takahara (1995) have found that $f=1.7$ is a good
approximation for accretion parameters appropriate for an X-ray binary.

Nevertheless, adding a soft X-ray component ({\scriptsize{DISKBB}} in XSPEC) to the power law 
only in four cases the reduced chi squares decrease below two, with
strong residuals at energy $>$40keV, indicating that a high energy tail
is also needed to fit these data. X-ray spectra clearly show the necessity of a soft X-ray 
component that is well described by 
a multicolor disk blackbody and of a high-energy curvature that is well reproduced by 
a cutoff power law. 
These fit results of the broad band spectrum using the models with
two emission components
are reported
in Table 3, available at the CDS. 
We also tried to fit this data set with a
model consisting of a only one Comptonization component.
For this hard component we used the model {\scriptsize{COMPTT}} in {\scriptsize{XSPEC}} (Titarchuk 1994),
assuming a spherical geometry
for the Comptonizing region and
leaving free the following parameters: the temperature of the Comptonizing
electrons kT${\rm _e}$, the plasma optical depth
${\rm \tau_p}$ and the input
temperature of the soft photon Wien distribution kT${\rm _0}$.
Using this model we can estimate the sizes of the emitting seed photons region,
assuming their emission as blackbody with temperature T$_0$
and radius R$_{seed}$ (in 't Zand et al. 1999).
We obtain $R_{\rm seed}\simeq 8.3\,{\rm km}\,
(L_{\rm C}/10^{37}\,{\rm erg\, s}^{-1})^{1/2} (1+y)^{-1/2} (kT_0/1\,{\rm keV})^{-2}$,
where $L_{\rm C}$ is the luminosity of the Comptonization component.
We have estimated the amplification factor of the seed luminosity
by Comptonization as $(1+y)$, where $y$ is the Comptonization parameter,
$y=4kT_{\rm e}\max(\tau,\tau^2)/m_{\rm e} c^2$.
However, this model alone is not representative of the entire sample, indeed
8 fits out 13 require a soft component,  
assuming a significative improvement of the fit for F-test chance probabilities $<5\times10^{-3}$.
Fit results of the broad band spectrum using the models with both
{\scriptsize{DISKBB}} and {\scriptsize{COMPTT}}  components
are reported
in Table 4, available at the CDS, while spectra and residuals with respect to the corresponding overplotted best fits
are shown in figure 1, available at the CDS. 
We have not find differences in the spectral results using the
simple black body model instead of the multi color disk black body.
Unfortunately the 
available data do not allow us to discriminate
the emission from the neutron star and the accretion disk.
% Figure 1 available electronically only

\begin{figure*}
\centering
\begin{minipage}[b]{16cm}
%{\em{Left:} 4U~1820-30; \em{Right:}}
%\hspace{1.0cm}
   {\includegraphics[height=7cm, width=3.2cm, angle=-90]{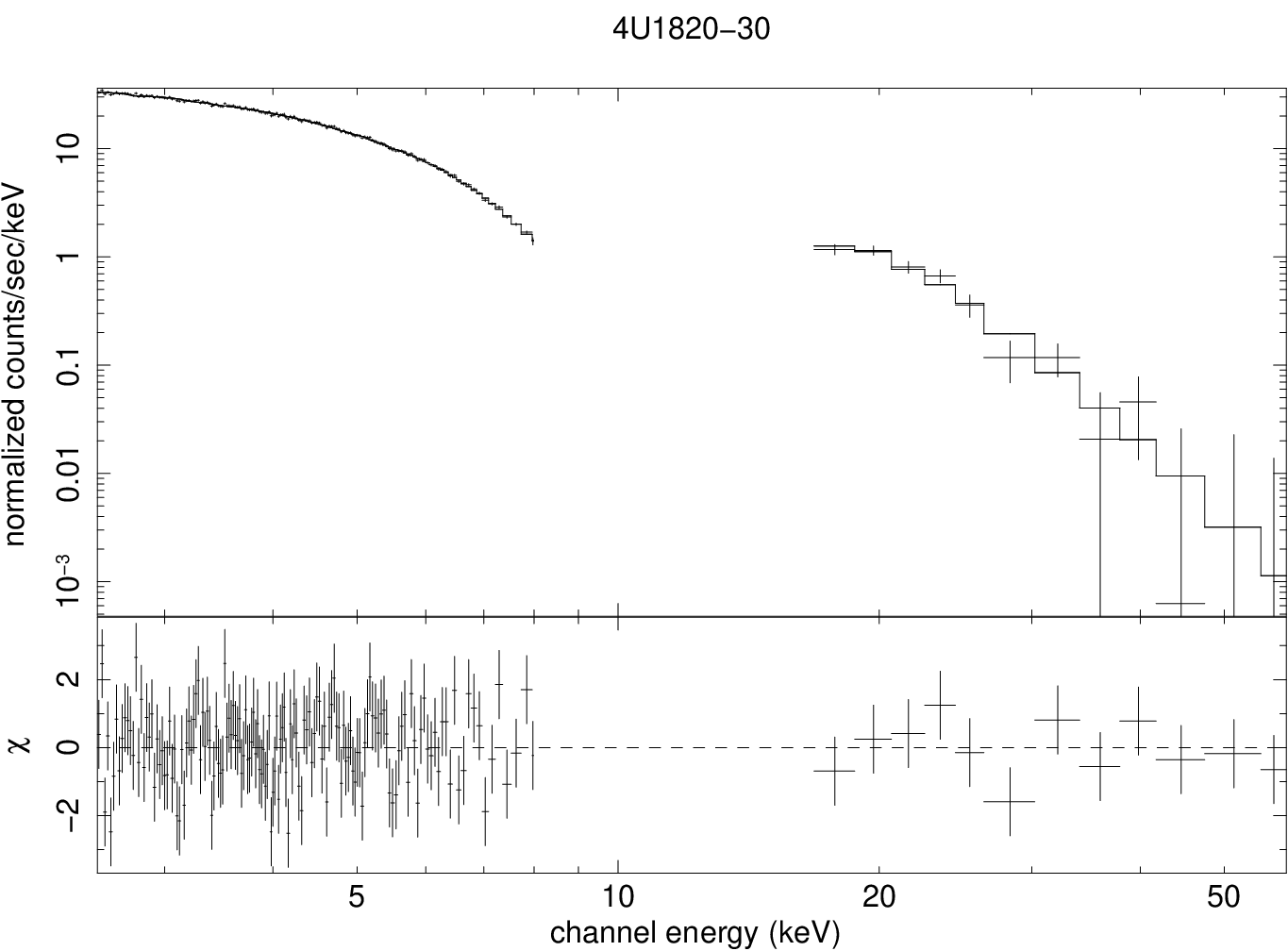}}
\vspace{1.5mm}
%\hspace{0.6cm}
   {\includegraphics[height=7cm, width=3.2cm, angle=-90]{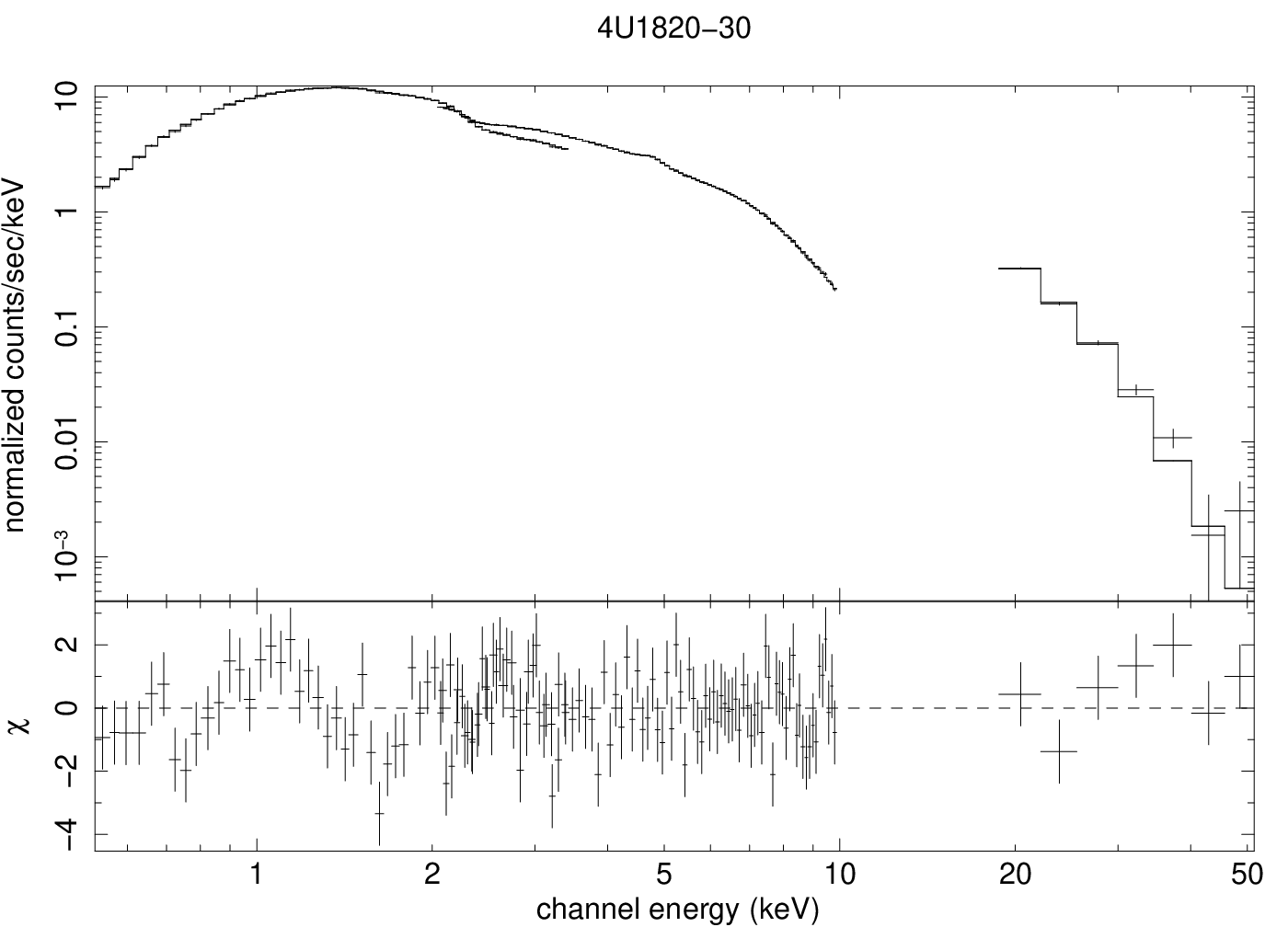}}
\vspace{1.5mm}
%\hspace{1.0cm}
   {\includegraphics[height=7cm, width=3.2cm, angle=-90]{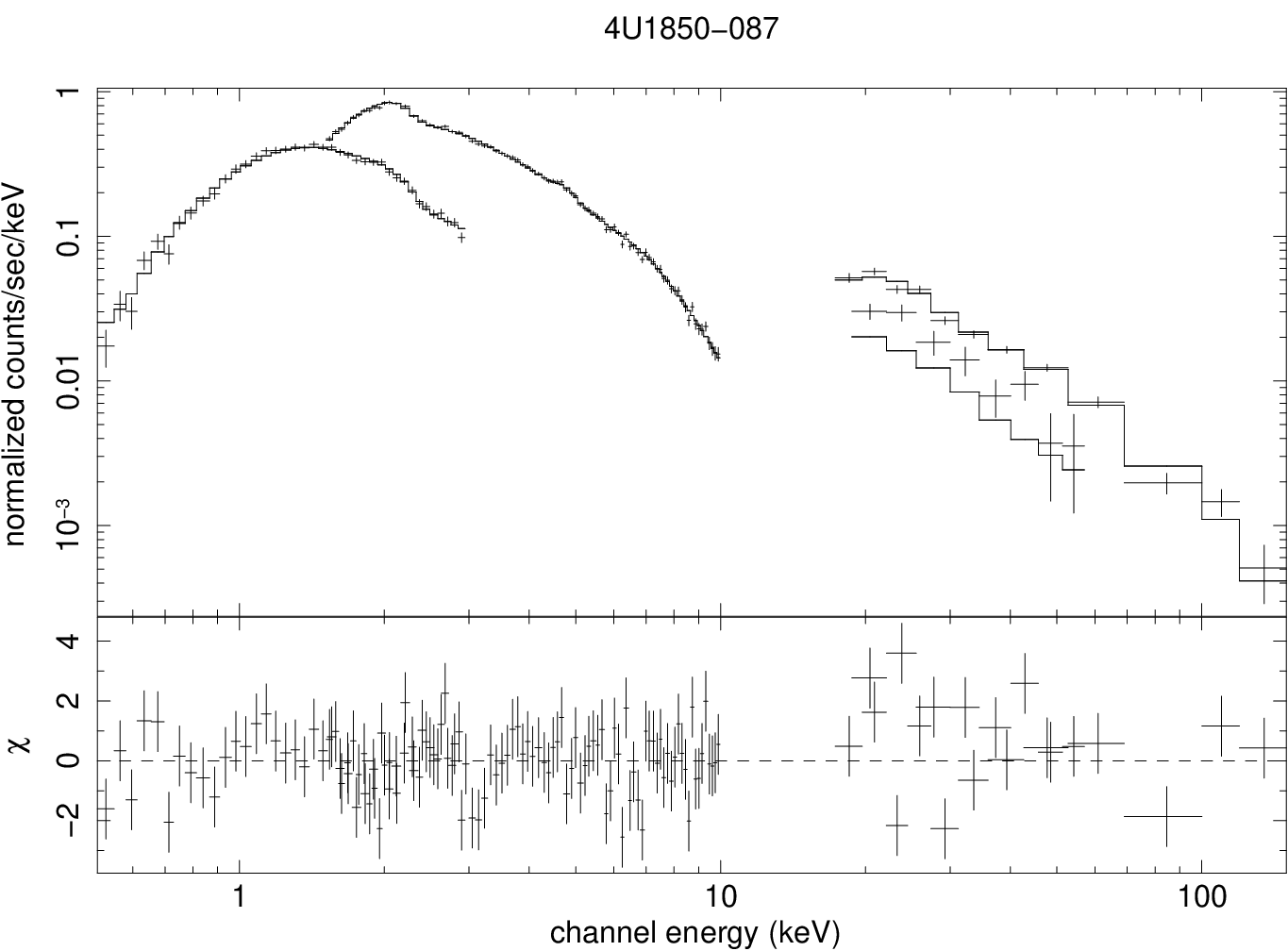}}
\vspace{1.5mm}
%\hspace{0.6cm}
   {\includegraphics[height=7cm, width=3.2cm, angle=-90]{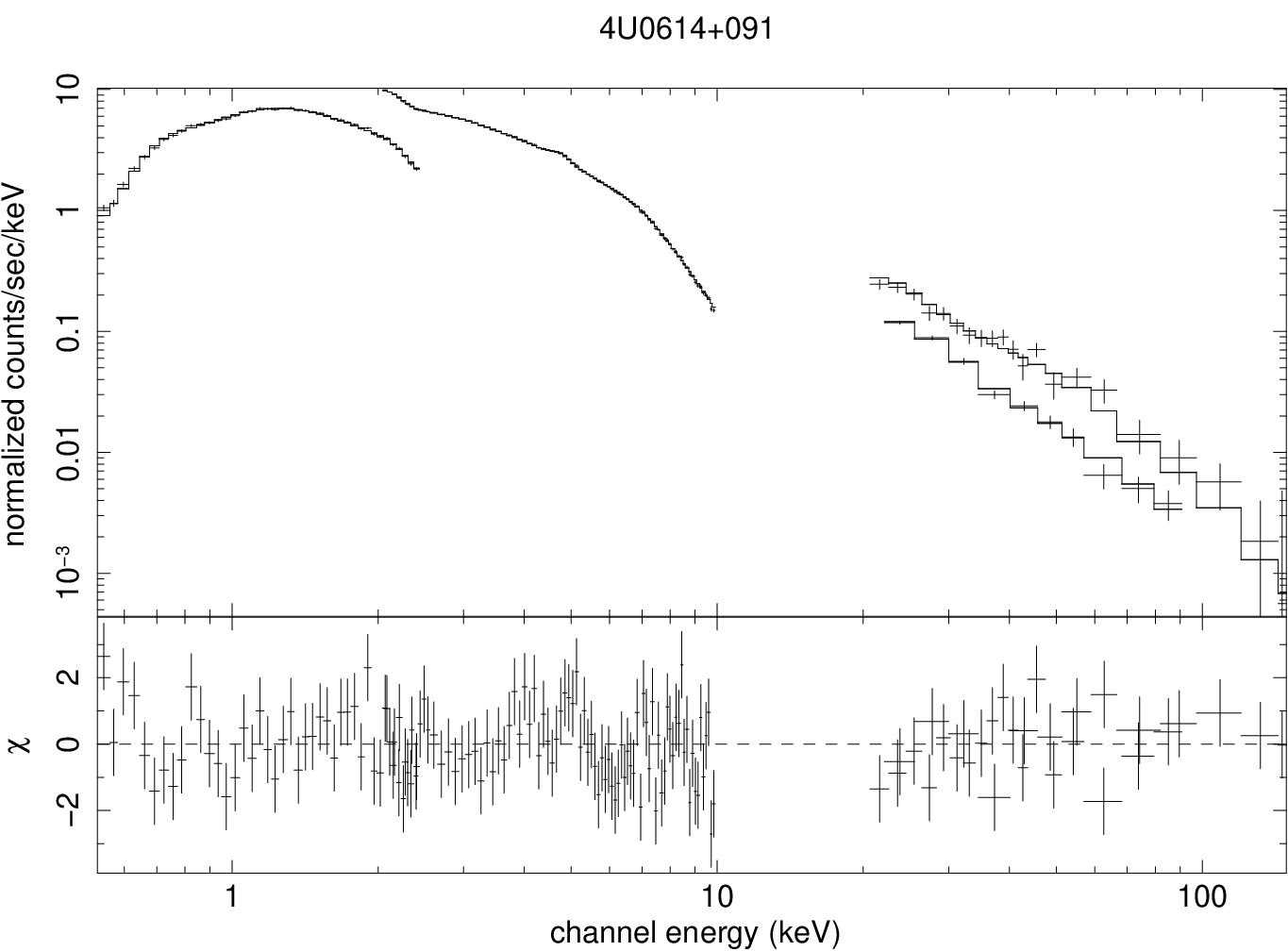}}
\vspace{1.5mm}
%\hspace{1.0cm}
   {\includegraphics[height=7cm, width=3.2cm, angle=-90]{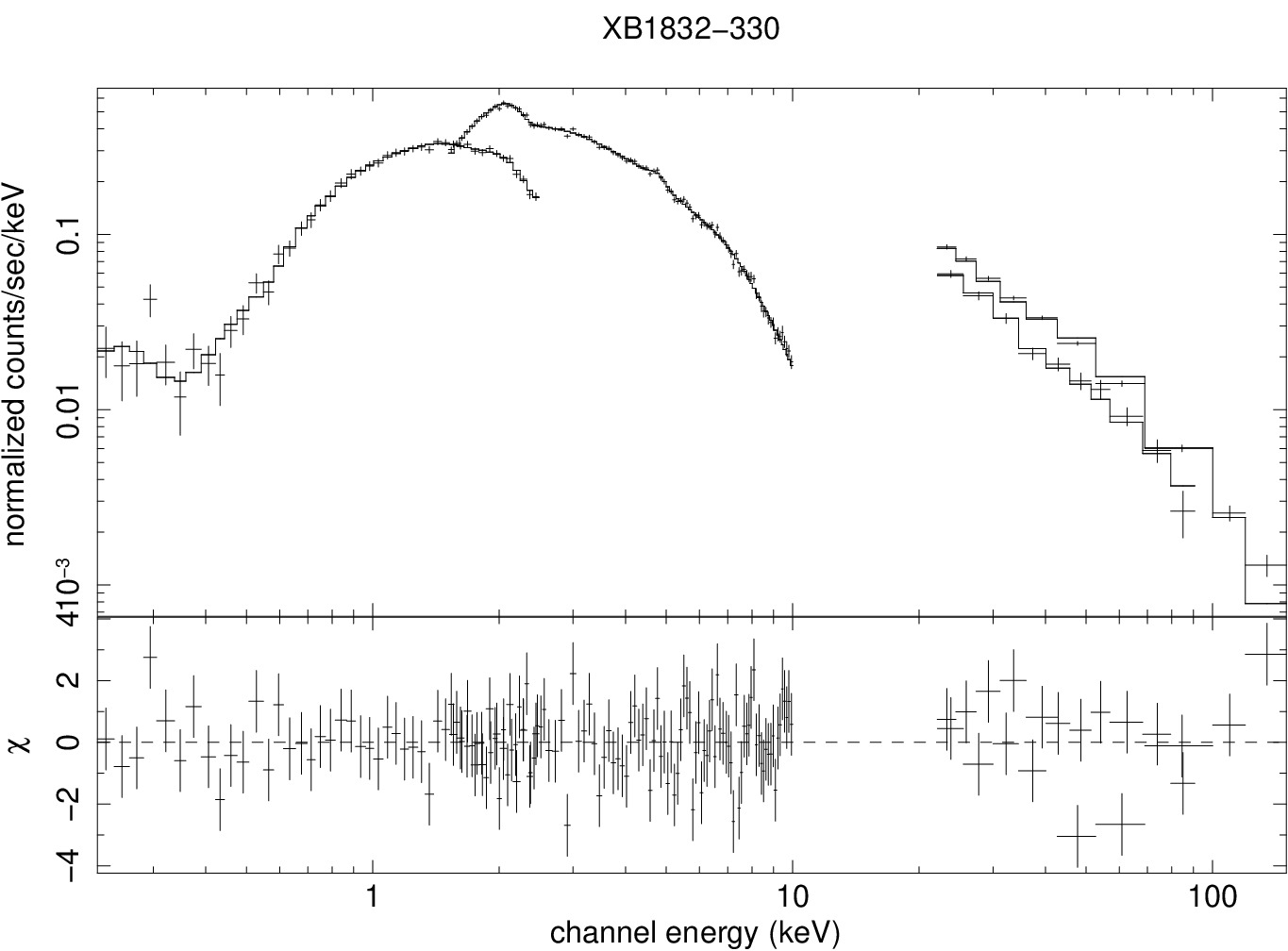}}
\vspace{1.5mm}
%\hspace{0.6cm}
   {\includegraphics[height=7cm, width=3.2cm, angle=-90]{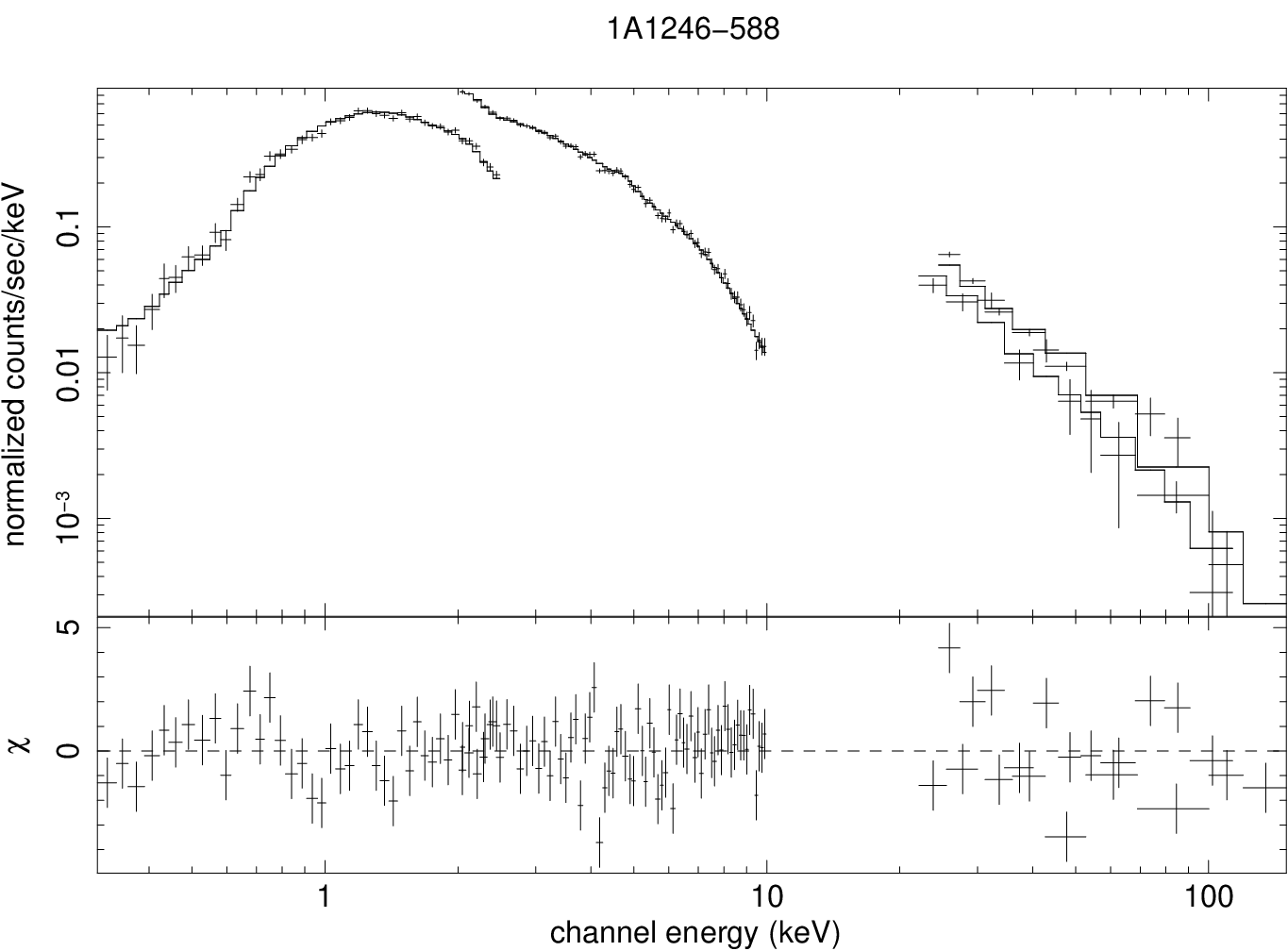}}
\vspace{1.5mm}
%\hspace{1.0cm}
   {\includegraphics[height=7cm, width=3.2cm, angle=-90]{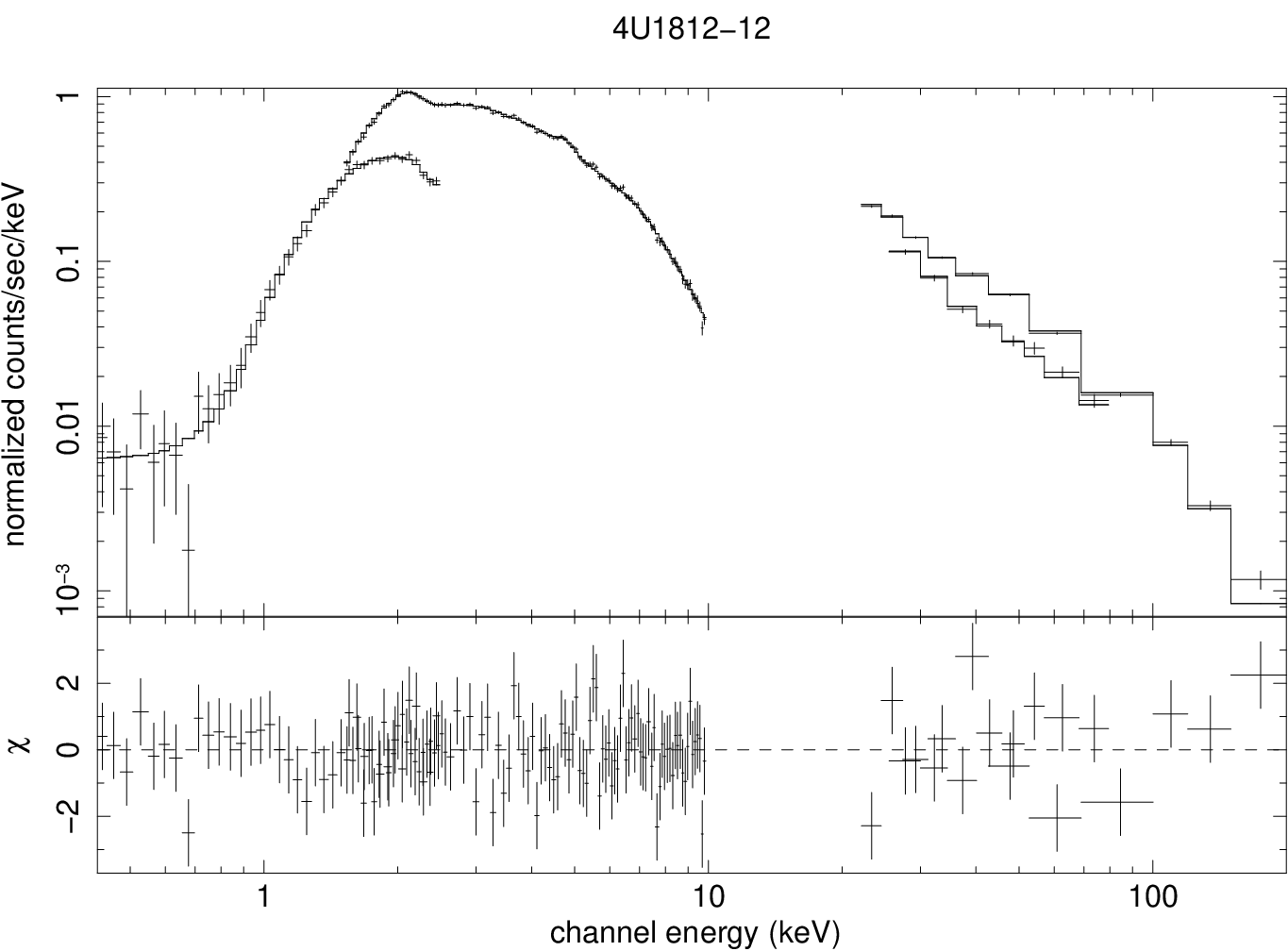}}
\vspace{1.5mm}
%\hspace{0.6cm}
  {\includegraphics[height=7cm, width=3.2cm, angle=-90]{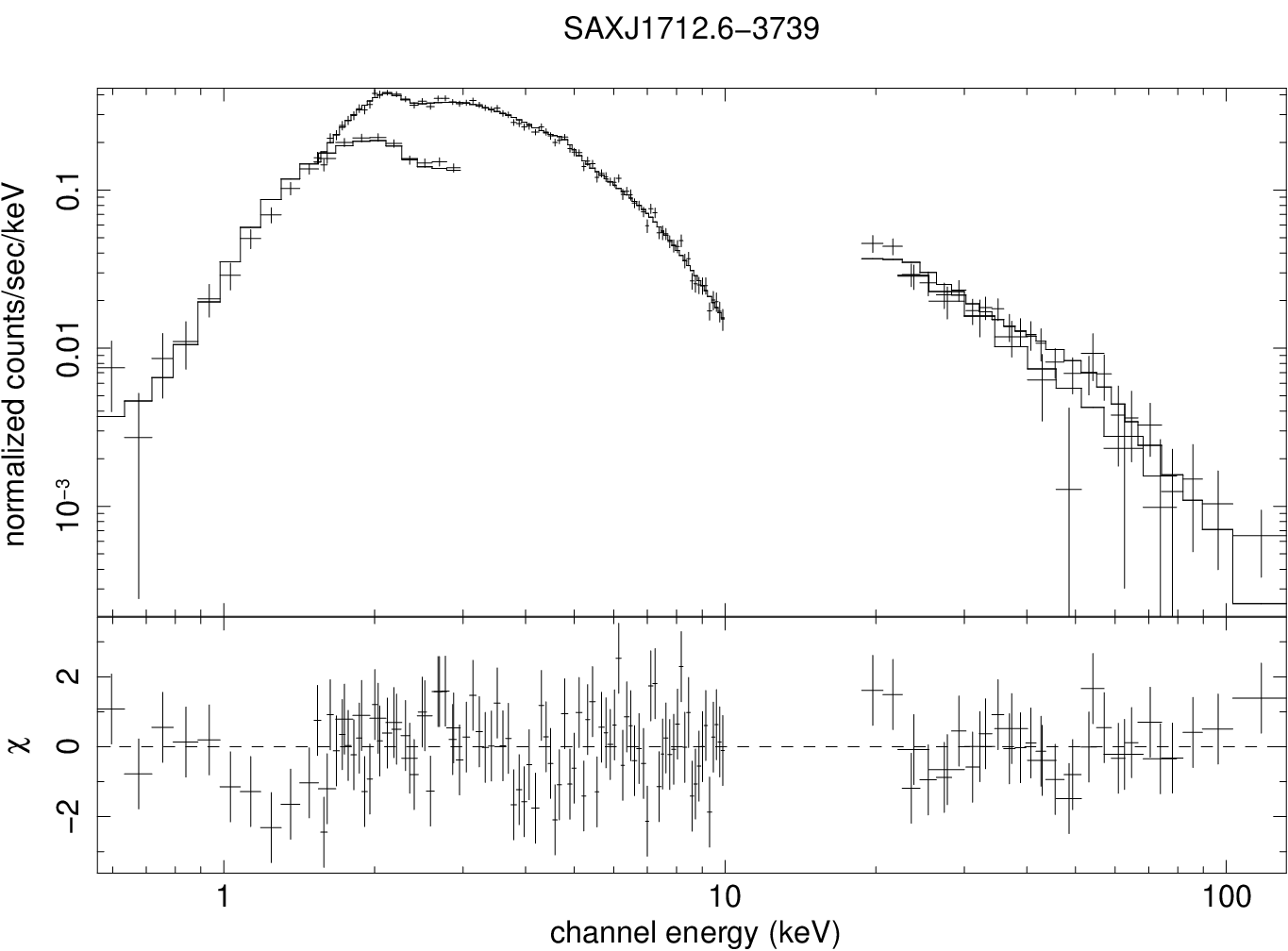}}
\vspace{1.5mm}
%\hspace{1.0cm}
   {\includegraphics[height=7cm, width=3.2cm, angle=-90]{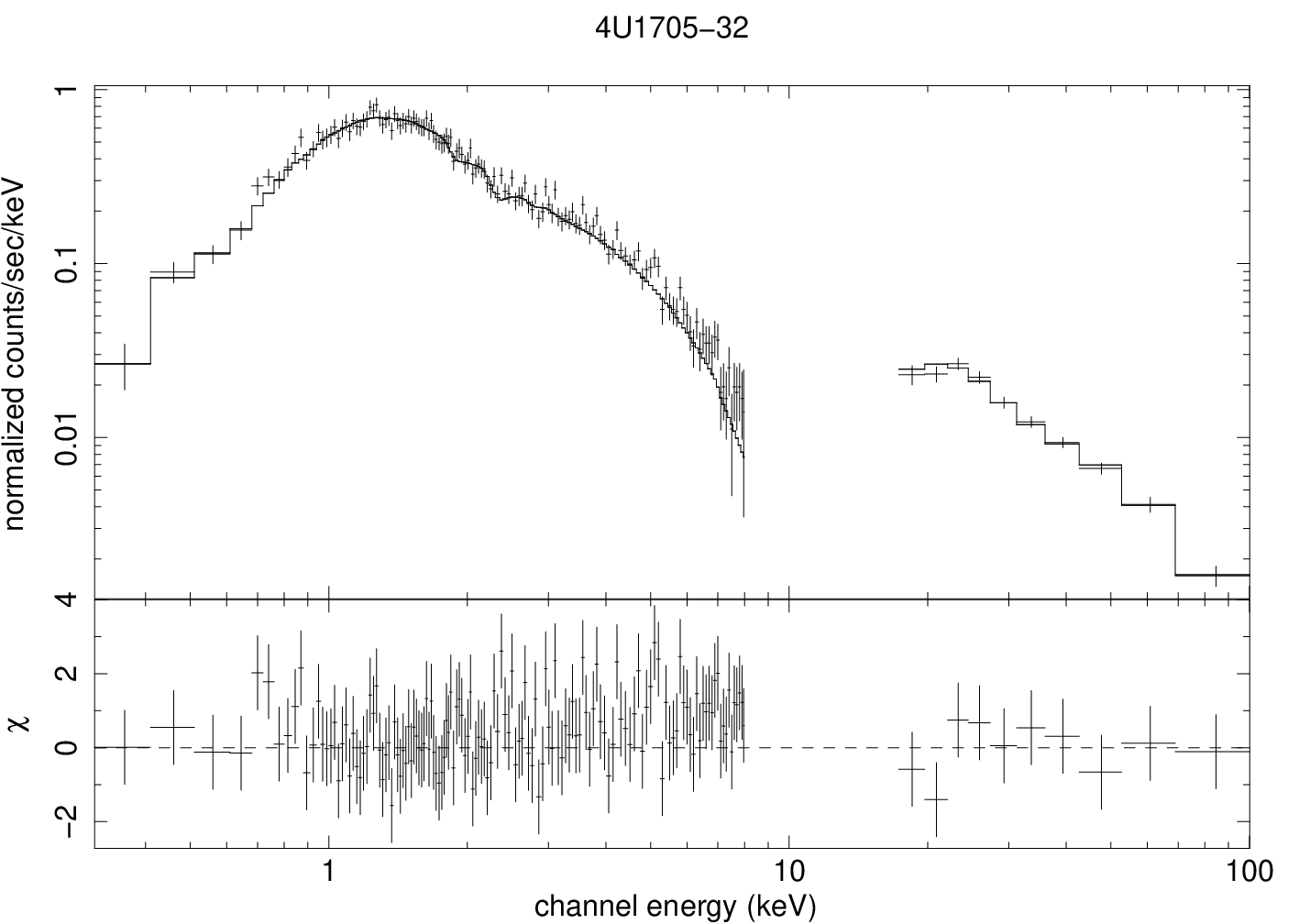}}
\vspace{1.5mm}
%\hspace{0.6cm}
   {\includegraphics[height=7cm, width=3.2cm, angle=-90]{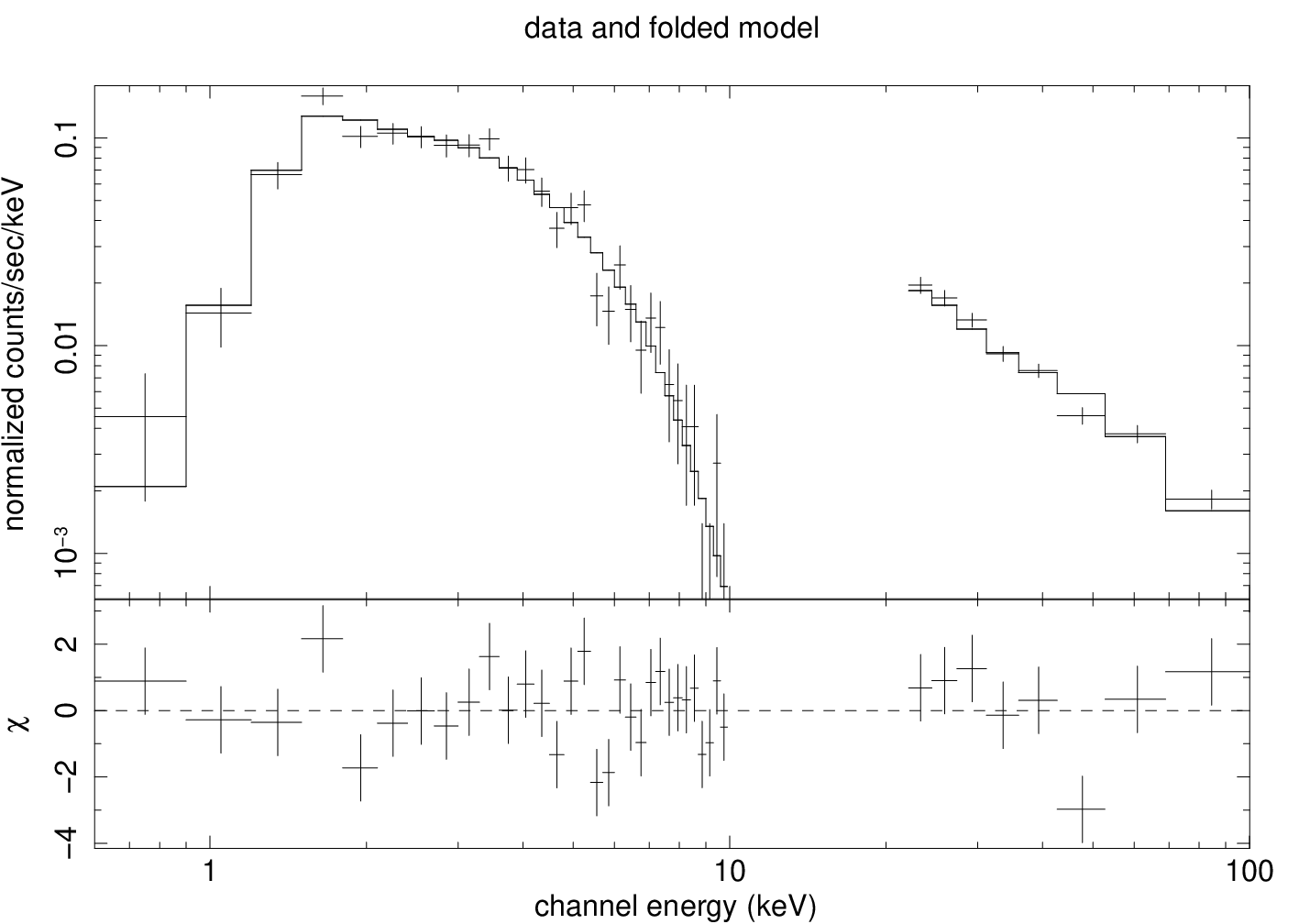}}
\vspace{1.5mm}
%\hspace{1.0cm}
   {\includegraphics[height=7cm, width=3.2cm, angle=-90]{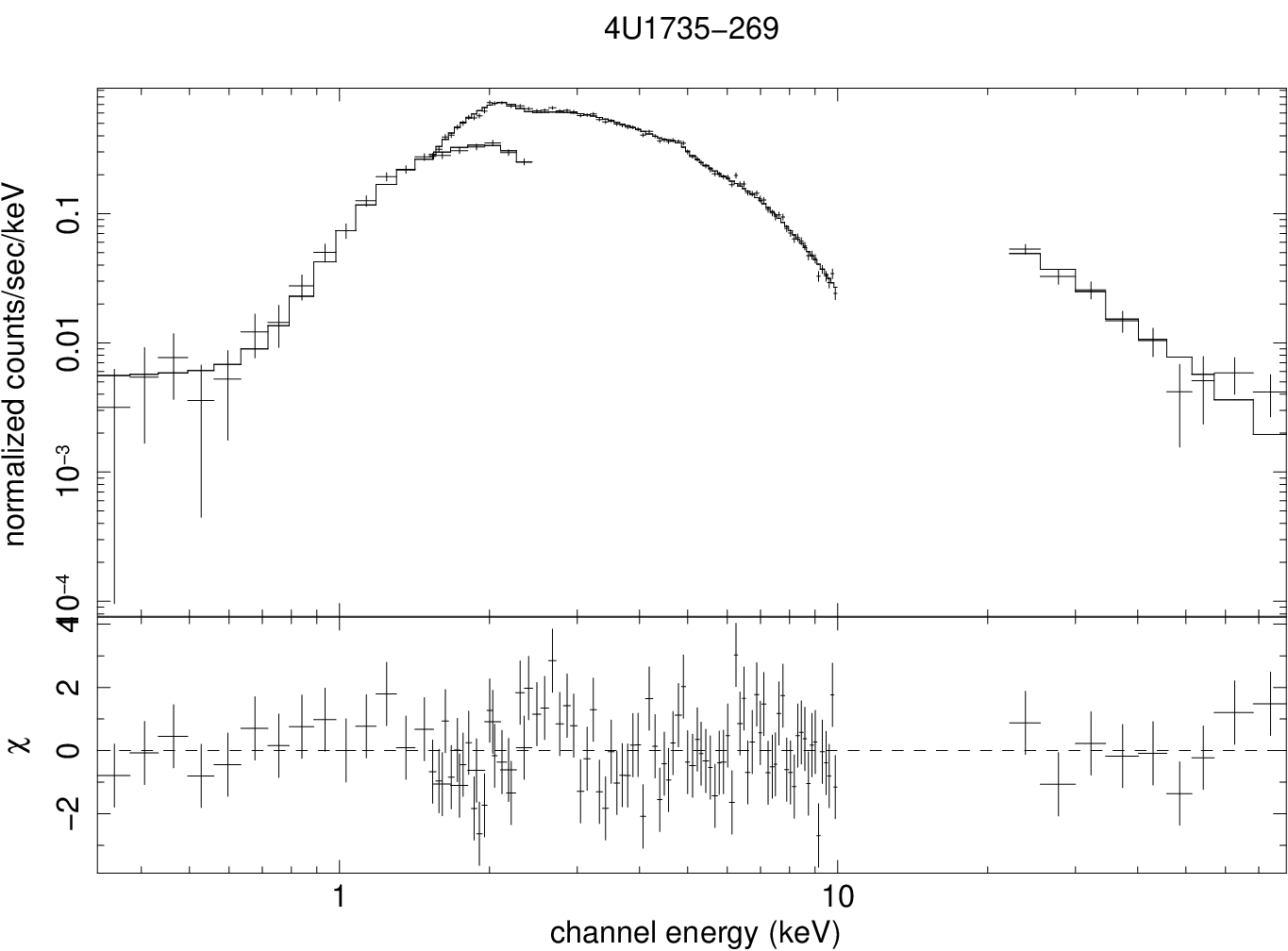}}
\vspace{1.5mm}
%\hspace{0.6cm}
   {\includegraphics[height=7cm, width=3.2cm, angle=-90]{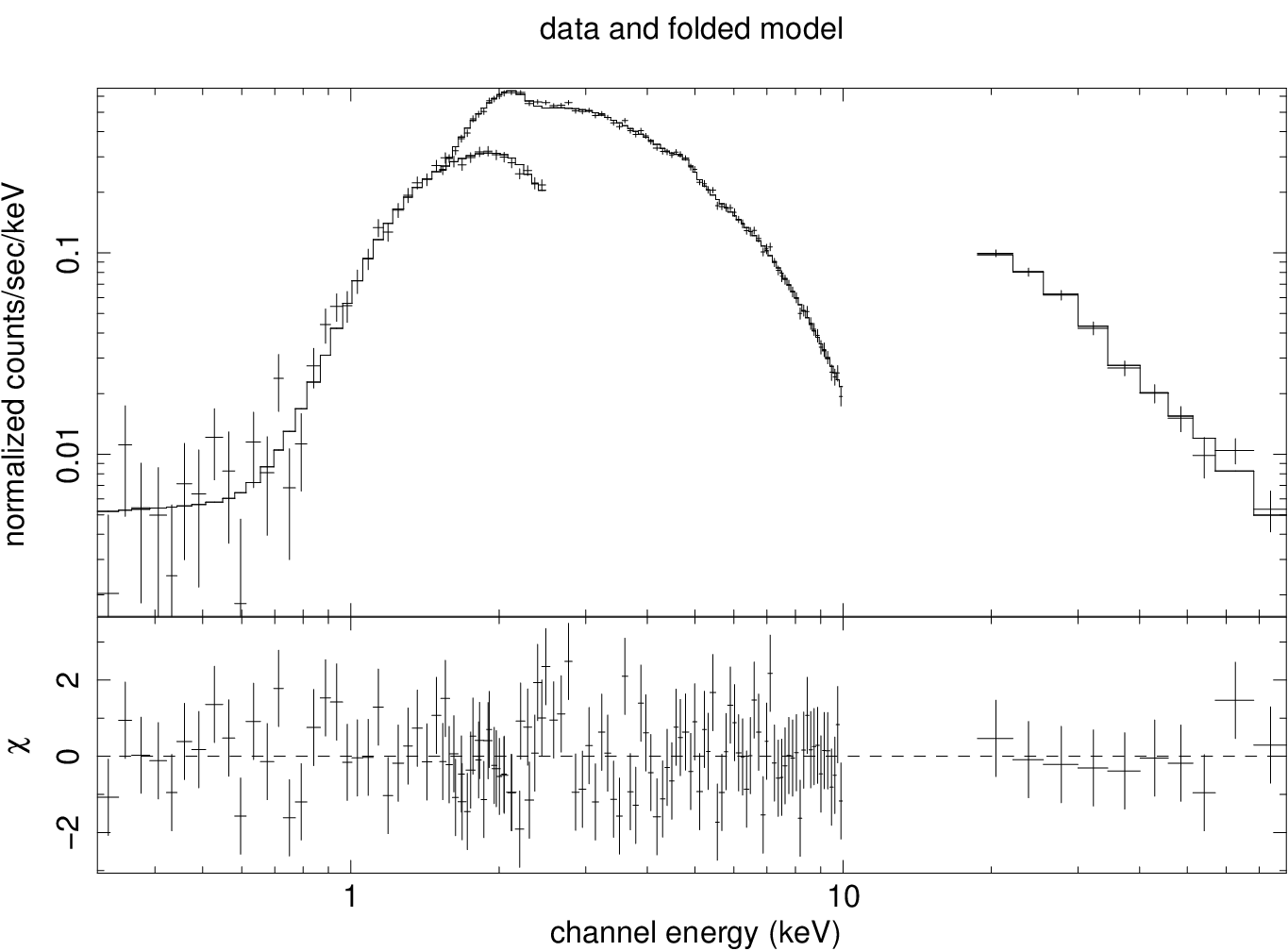}}
\vspace{1.5mm}
%\hspace{1.0cm}
   {\includegraphics[height=7cm, width=3.2cm, angle=-90]{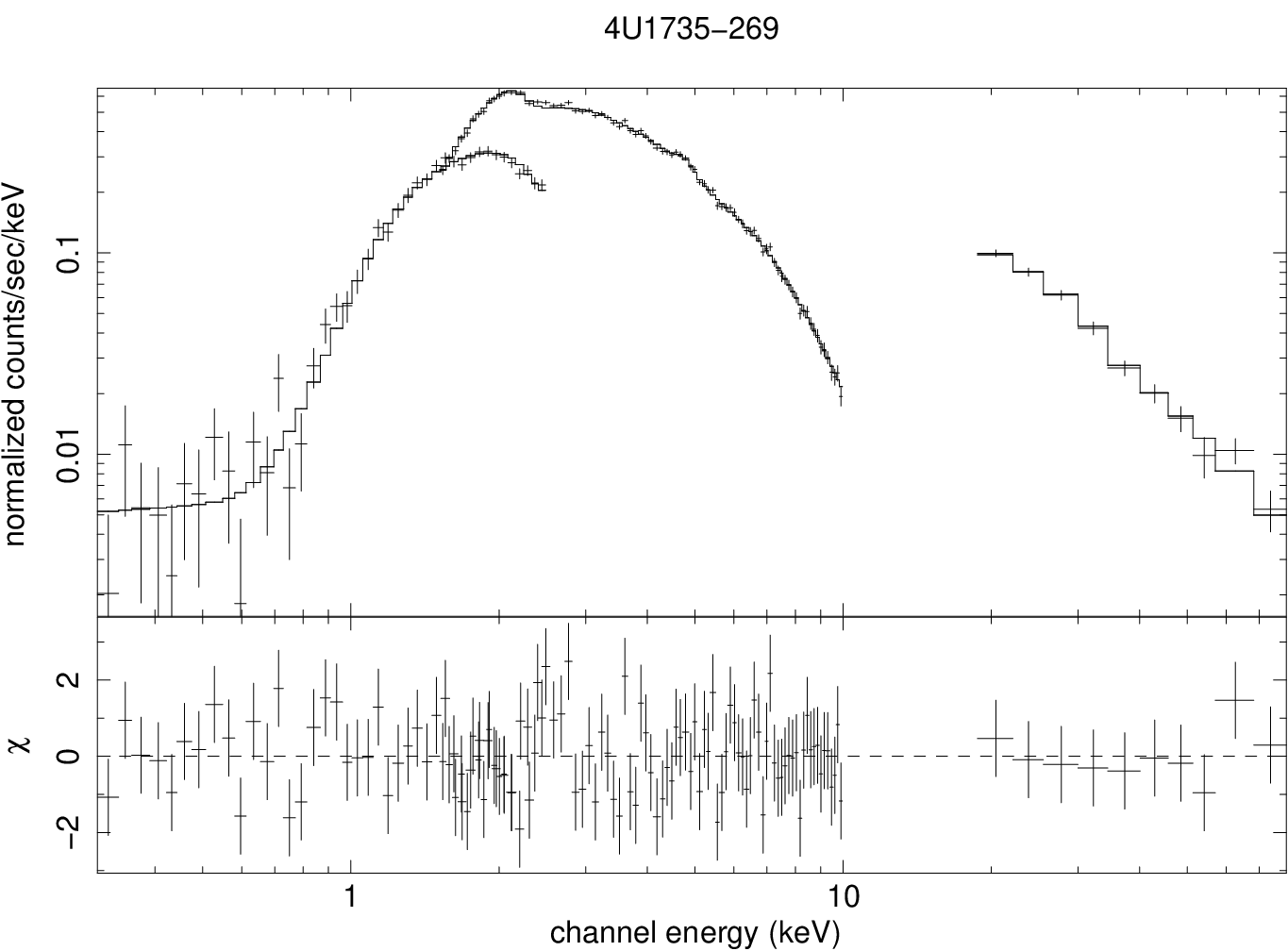}}
\caption{Spectra of burster UCXB sample and the
residuals with respect to the corresponding overplotted best fits, using a model consisting of {\scriptsize{DISKBB}} and 
{\scriptsize{COMPTT}} components.}
\label{count2}
\end{minipage}
\end{figure*}

% Table 3 available electronically only

\begin{table*}
\caption[]{BeppoSAX spectral results using the cutoff powerlaw and
disk-blackbody model.
L is the 0.1--100 keV (assumed isotropic, in units of 10$^{36}$~erg~s$^{-1}$)
source unabsorbed luminosity using the average values of distances given in
Table~\ref{log}.
The units of ${\rm N_H}$ are
$10^{22}$~atom~cm$^{-2}$.
$\theta$ is the inclination angle of the disk.
All uncertainties are quoted at 90\%
confidence}
%\begin{flushleft}
%\scriptsize
\begin{tabular}{llllllll}
\hline\noalign{\smallskip}
\\
\multicolumn{8}{c}{Continuum parameters}
\\
Source & \mc{1}{c}{L} &  \mc{1}{c}{${\rm N_H}$}
&\mc{1}{c}{${\Gamma}$} & \mc{1}{c}{kT${\rm _{cut}}$}
& \mc{1}{c}{${\rm kT_{eff}}$}
& \mc{1}{c}{${\rm r_{eff}} (\cos \theta)^{0.5}$}  & $\chi^2$/dof \\
  &$10^{36}erg s^{-1}$ & & & \mc{1}{c}{(keV)}  &\mc{1}{c}{(keV)} &(km) & \\
%\noalign{\smallskip\hrule\smallskip}
\\
\hline
\\
1) 4U 1820-303$^a$         &$68^{+47}_{-26}$      &$<0.58$                &$1.1^{+1.0}_{-0.8}$    &$6^{+5}_{-2}$      &$1.6^{+0.5}_{-0.2}$   &$6\pm1$           &598/556 \\
\\
2) 4U 1820-303$^a$                        &$29^{+4}_{-3}$        &$0.16\pm0.05$          &$0.3\pm0.1$            &$4.4^{+0.2}_{-0.4}$&$0.4\pm0.2$           &$12.0\pm0.6$         &179/120 \\
\\
4U 1850-087            &$2.2\pm0.3$           &$0.47\pm0.03$          &$2.03\pm0.08$          &$120^{+80}_{-92}$  &$0.37\pm0.02$         &$3.5\pm0.4$           &133/121\\
\\
4U 0614+091$^b$        &$5\pm1$               &$0.40\pm0.05$          &$2.36\pm0.06$          &$100^{+19}_{-6}$   &$0.15\pm0.02$         &$16^{+6}_{-3}$        &260/155 \\
\\
XB 1832-330            &$8\pm1$               &$0.26\pm0.02$          &$1.54^{+0.04}_{-0.05}$ &$62\pm8$           &$0.68\pm0.04$         &$1.0\pm0.2$            &209/137 \\
\\
1A 1246-588            &$4^{+1}_{-3}$         &$0.5\pm0.1$            &$2.08\pm0.01$          &$61\pm14$          &$0.10\pm0.02$         &$<352^{+356}_{-123}$   &940/650 \\
\\
4U 1812-12             &$3.2\pm0.6$           &$1.78\pm0.08$          &$1.89\pm0.05$          &$136\pm28$         &$0.38\pm0.05$         &$2.4^{+1.3}_{-0.5}$    &633/890 \\
\\
SAX J1712.6-3739$^{c}$ &$1.3\pm0.3$           &$1.4\pm0.1$            &$1.4\pm0.1$            &$30\pm6$           &$0.40\pm0.04$         &$2.8\pm0.3$           &138/133  \\
\\
4U 1705-32             &$4\pm2$               &$0.37\pm0.05$          &$1.8\pm0.3$            &$>50$              &$0.7\pm0.3$           &$1.0\pm0.4$           &148/148 \\
\\
1RXS J172525.5-32      &$<4.7$           &$2.1^{+0.4}_{-0.3}$    &$1.7\pm0.3$            &$119^{+465}_{-59}$ &$0.5\pm0.2$           &$<9$                  &48/32 \\
\\
1) SLX 1735-269           &$4\pm2$               &$1.2\pm0.1$            &$1.3\pm0.3$            &$28^{+18}_{-10}$   &$0.60\pm0.05$         &$1.8^{+0.3}_{-0.2}$   &135/102 \\
\\
2) SLX 1735-269                       &$4^{+2}_{-1}$         &$1.2\pm0.1$            &$1.5\pm0.2$            &$42^{+22}_{-10}$   &$0.58\pm0.03$         &$1.7\pm0.2$           &119/124\\
\\
SLX 1737-282           &$1.0^{+0.4}_{-0.8}$   &$1.8\pm0.3$            &$1.6\pm0.2$            &$61_{-20}^{+54}$   &$0.47\pm0.06$         &$1.2_{-0.2}^{+0.6}$        &119/94 \\
\\
\hline\hline
\\
\multicolumn{8}{c}{Emission Lines parameters}
\\
\\
Name&\multicolumn{2}{c}{E$_{line}$}&\multicolumn{2}{c}{$\sigma$}&\multicolumn{2}{c}{flux}\\
&\multicolumn{2}{c}{keV}&\multicolumn{2}{c}{keV}&\multicolumn{2}{c}{$10^{-3} ph cm^{-2}s^{-1}$}\\
\\
\hline
\\
1) 4U1820-303&\multicolumn{2}{c}{$6.1\pm0.2$}&\multicolumn{2}{c}{$0.5\pm0.3$}&\multicolumn{2}{c}{$6\pm4$}\\
\\
2) 4U1820-303&\multicolumn{2}{c}{$6.6\pm0.3$}&\multicolumn{2}{c}{$1.0\pm0.2$}&\multicolumn{2}{c}{$2_{-1}^{+4}$}\\
\\
4U0614+091&\multicolumn{2}{c}{$0.71\pm0.03$}&\multicolumn{2}{c}{$<0.7$}&\multicolumn{2}{c}{$232_{-48}^{+68}$}\\
\\
4U0614+091&\multicolumn{2}{c}{$6.0\pm0.1$}&\multicolumn{2}{c}{$1.4\pm0.3$}&\multicolumn{2}{c}{$6\pm2$}\\
\\
SLX1737-282$^{a}$&\multicolumn{2}{c}{$6.6\pm0.3$}&\multicolumn{2}{c}{$0.4_{-0.3}^{+0.4}$}&\multicolumn{2}{c}{$0.08_{-0.04}^{+0.08}$}\\
\\
\hline
%\noalign{\smallskip\hrule\smallskip}
\end{tabular}\\
$^{a}$A Gaussian component is required by the data\\
$^{b}$Two Gaussian components are required by the data\\
$^{c}$ fit results from Fiocchi et al. 2008\\
%$^{1a}$ fit results from Sidoli et al. 2001\\
%$^{1b}$ fit results from Sidoli et al. 2007\\
%$^{2}$ fit results from ???????????? al. ????\\
%$^{3}$ fit results from in't Zand et al. 2005\\
%$^{4}$ fit results from Guainazzi et al. 1998\\
%\end{flushleft}
\label{fit1}
\end{table*}

\newpage
\begin{table*}
\caption[]{BeppoSAX spectral results using the standard {\sc comptt} and
disk-blackbody model.
L is the 0.1--100 keV (assumed isotropic, in units of 10$^{36}$~erg~s$^{-1}$)
source unabsorbed luminosity using the average values of distances given in
Table~\ref{log}.
The units of ${\rm N_H}$ and $\mdot\/$ are
$10^{22}$~atom~cm$^{-2}$ and $10^{-10}$~M$_{solar}$/yr, respectively.
$\theta$ is the inclination angle of the disk.
All uncertainties are quoted at 90\%
confidence}
%\begin{flushleft}
\scriptsize
\begin{tabular}{llllllllccc}
\hline\noalign{\smallskip}
\\
\multicolumn{11}{c}{Continuum parameters}
\\
Source & \mc{1}{c}{L} &  \mc{1}{c}{${\rm N_H}$}
&\mc{1}{c}{k${\rm T_0}$} & \mc{1}{c}{kT${\rm _e}$}
& \mc{1}{c}{${\rm \tau_p}$} & \mc{1}{c}{${\rm kT_{eff}}$}
& \mc{1}{c}{${\rm r_{eff}} (\cos \theta)^{0.5}$} & \mc{1}{c}{R$_{comptt}$}  & $\mdot$ &$\chi^2$/dof \\
  &$10^{36}erg s^{-1}$ & & \mc{1}{c}{(keV)} & \mc{1}{c}{(keV)} & &\mc{1}{c}{(keV)} & \mc{1}{c}{(km)} &km && \\
%\noalign{\smallskip\hrule\smallskip}
\\
\hline
\\
1) 4U 1820-303$^a$&$59^{+40}_{-15}$ &$0.11\pm0.08$ &$0.2\pm0.1$
&$3.3\pm0.3$ &$15\pm3$
 &$1.2\pm0.2$ &$6.7\pm2.3$&140  &104&599.3/555 \\
\\
2) 4U 1820-303$^a$&$33\pm3$ &$0.14^{+0.07}_{-0.02}$ &$0.23^{+0.03}_{-0.22}$
&$3.17\pm0.09$ &$16.6\pm0.7$
 &$0.56^{+0.01}_{-0.05}$ & $19.1^{+0.9}_{-1.1}$&102   &58&167.3/119 \\
\\
4U 1850-087&$2.0^{+3.5}_{-0.4}$ &$0.33^{+0.06}_{-0.09} $ &$<$0.3 &$38^{+54}_{-11}$ &$2.1^{+1.9}_{-1.5}$ &$0.39\pm0.03$ &$9.5^{+2.0}_{-1.5}$ &25 &3.5&133.0/120\\
\\
4U 0614+091$^b$&$3\pm1$&$0.24\pm0.01$ &$0.78^{+0.03}_{-0.02}$ &$141^{+39}_{-71}$ &$0.17^{+0.05}_{-0.07}$ &$0.28\pm0.01$ &$35\pm3$ &7 &5.3&216.3/154 \\
\\
XB 1832-330&$7\pm1$&$0.03^{+0.02}_{-0.01}$ &$0.44^{+0.03}_{-0.01}$ &$23^{+3}_{-1}$&$4.1^{+0.2}_{-0.3}$ &$0.12^{+0.05}_{-0.03}$ &$23^{+54}_{-9}$ &18 &12&165.7/136 \\
\\
1A 1246-588&$3\pm1$ &$0.49\pm0.03$ &$0.25\pm0.02$ &$19^{+6}_{-4}$ &$3.5^{+0.6}_{-1.3}$ &$0.10\pm0.01$ &$479^{+349}_{-133}$ &43&5.3 &888.8/649 \\
\\
4U 1812-12&$2.6^{+2.2}_{-0.4}$ &$1.4\pm0.1$ &$0.38^{+0.08}_{-0.06}$ &$60^{+12}_{-13}$ &$1.5\pm0.3$ &$0.29^{+0.06}_{-0.06}$ &$<29$ &21&4.6 &630/889 \\
\\
SAX J1712.6-3739$^{c}$ &$1.6^{+0.5}_{-1.2}$ &$1.4\pm0.3$ &$0.5\pm0.2$ &$24^{+22}_{-9}$ &$3\pm2$ &$0.4\pm0.1$ &$<32$ &10 &3.5&110/132  \\
\\
4U 1705-32&$3^{+7}_{-2}$ &$0.3\pm0.1$ &$0.6^{+0.3}_{-0.2}$ &$25^{+63}_{-3}$ &$3.5^{+1.5}_{-0.2}$ &$0.3\pm0.1$ &$17^{+26}_{-3}$ &7&5.3 &152/147 \\
\\
%%4U 1724-307&3-7 &1.13f &1.98f &$75^{+80}_{-52}$ &$0.8^{+1.8}_{-0.5}$ &... &... &... &3.0/5 \\
%%\\
1RXS J172525.5-32&$<4.7$&$2.0^{+0.3}_{-0.6}$ &$0.2^{+0.8}_{-0.1}$ &$67^{+45}_{-40}$ &$1.5\pm0.6$&$0.6^{+0.1}_{-0.4}$ &$<9$ &$<75$ &$8.3$ &43.9/31 \\
\\
1) SLX 1735-269&$3\pm2$ &$0.9^{+0.3}_{-0.2}$ &$0.52^{+0.09}_{-0.03}$&$15^{+32}_{-5}$ &$5^{+1}_{-3}$ &$0.4^{+0.4}_{-0.3}$ &$<$12 & 10&5.3 &133/101 \\
\\
2) SLX 1735-269    &$3\pm2$ &$1.0\pm0.2$ &$0.5^{+0.2}_{-0.1}$&$21^{+19}_{-6}$ &$4^{+1}_{-2}$ &$0.4^{+0.3}_{-0.1}$ &$<$12 & 9 &5.3&97/123 \\
\\
SLX 1737-282$^{a}$&$1.0^{+0.2}_{-0.5}$ &$1.4\pm0.2$ &$0.5\pm0.1$ &$30_{-12}^{+40}$ &$2.9_{-0.8}^{+1.6}$ &$0.4\pm0.1$ &$5_{-1}^{+17}$ &7&1.8 &117.6/93 \\
\\
%         & & & & & & & & & \\
\hline\hline
\\

\multicolumn{11}{c}{Emission Lines parameters}
\\
\\
&&Name&\multicolumn{2}{c}{E$_{line}$}&\multicolumn{2}{c}{$\sigma$}&\multicolumn{2}{c}{flux}&\\
&&&\multicolumn{2}{c}{keV}&\multicolumn{2}{c}{keV}&\multicolumn{2}{c}{$10^{-3} ph cm^{-2}s^{-1}$}&\\
\\
\hline
\\
&&1) 4U1820-303&\multicolumn{2}{c}{$6.0\pm0.2$}&\multicolumn{2}{c}{$0.6\pm0.3$}&\multicolumn{2}{c}{$7_{-3}^{+5}$}&\\
\\
&&2) 4U1820-303&\multicolumn{2}{c}{$6.0\pm0.1$}&\multicolumn{2}{c}{$1.6\pm0.2$}&\multicolumn{2}{c}{$18_{-3}^{+5}$}&\\
\\
&&4U0614+091&\multicolumn{2}{c}{$0.72\pm0.03$}&\multicolumn{2}{c}{$<0.7$}&\multicolumn{2}{c}{$113_{-28}^{+78}$}&\\
\\
&&4U0614+091&\multicolumn{2}{c}{$6.4\pm0.1$}&\multicolumn{2}{c}{$1.4\pm0.2$}&\multicolumn{2}{c}{$6_{-1}^{+2}$}&\\
\\
&&SLX1737-282&\multicolumn{2}{c}{$6.6\pm0.3$}&\multicolumn{2}{c}{$0.4_{-0.3}^{+0.5}$}&\multicolumn{2}{c}{$0.09_{-0.04}^{+0.08}$}&\\
\\
\hline
%\noalign{\smallskip\hrule\smallskip}
\end{tabular}\\
$^{a}$A Gaussian component is required by the data\\
$^{b}$Two Gaussian components are required by the data\\
$^{c}$ Fit results from Fiocchi et al. 2008\\
%$^{1a}$ fit results from Sidoli et al. 2001\\
%$^{1b}$ fit results from Sidoli et al. 2007\\
%$^{2}$ fit results from ???????????? al. ????\\
%$^{3}$ fit results from in't Zand et al. 2005\\
%$^{4}$ fit results from Guainazzi et al. 1998\\
%\end{flushleft}
\label{fit2}
\end{table*}

When statistically required, a Gaussian component was added to the model
to reproduce the $K_{\alpha}$ iron line. 
We found that three sources only display meaningful emission lines.

Finally, we also tried to fit data adding to the model a partial covering fraction absorption.
Only in the case of the burster XB~1832-330 this component is statistically required by the fit
as previously reported by Parmar et al. 2001
(for details see notes on the individual sources below).

\begin{figure*}[h]
\begin{center}
\hspace{0.5cm}
{\includegraphics[height=9.0cm, angle=-90]{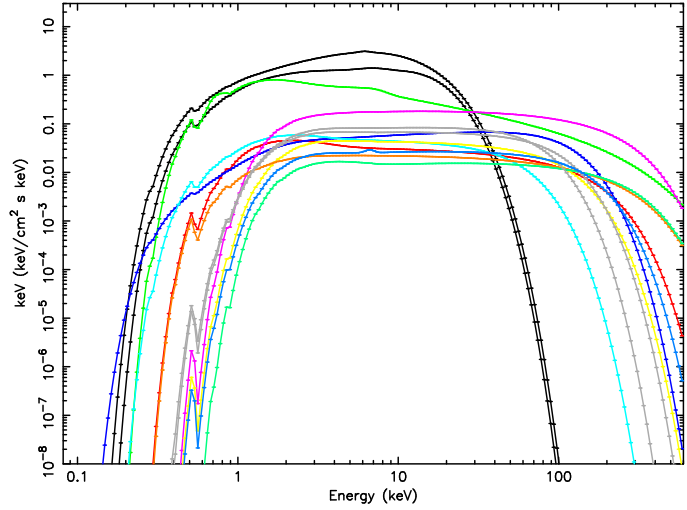}}
\caption{Comparison of the extrapolated models for the burster UCXB observations, using the Comptonization models.}
\label{model}
\end{center}
\end{figure*}

Hereafter, we add these peculiarities on the spectral fitting of individual sources:\\
{\em {1A1246-588}}:
this X-ray burster was detected with Swift on 2006 August 11
during an outburst. The authors suggested that it is a long Type-I 
X-ray burst with duration of at least 10 minutes
(Kong 2006). Such outburst was excluded during the extraction of the spectrum from 
the persistent emission, nevertheless the XRT/IBIS intercalibration 
constant is $\sim$4.\\
{\em {4U1735-269}}:
the light curve reported in this paper and previous analysis performed by 
Molkov et al. 2005 show spectral variations during \integral\/ observations in 2003,
for this reason here we report only \sax\/ data. \\
{\em {SLX 1737-282}}: we exclude PDS data because they are contaminated by  the near new \integral\/ 
source IGR~J17407-2808 (see Bird et al. 2007).\\
{\em {XB 1832-330}}: Figure 1, available at the CDS, shows that there are structured residuals remaining
in the low energy part of the spectrum (LECS and MECS data). 
Following indications in Parmar et al. 2001, 
we fitted data using a partial covering model.
This component is statistically highly significant, 
reducing $\chi^{2}/$d.o.f.\ from 209/137 to 154/135 for the model consisting of 
a disk black body plus a cutoff power law and from
166/136 to 142/134 for the model consisting of 
a disk black body plus a {\scriptsize{COMPTT}}, with the corresponding F-test chance
probabilities of $3\times 10^{-5}$ and $1\times 10^{-9}$, respectively.
Obtained spectral parameters using a partially covered disk-black body and COMPTT model
are the following: the fraction of the flux that undergoes extra absorption 
$f_{PCF}=0.2\pm0.1$,
$N_{PCF}$=$5.9\pm0.8\times10^{22}$atoms~cm$^{-2}$,
N$_H$=$4.5^{+2.6}_{-1.6}\times10^{20}$atoms~cm$^{-2}$, 
$kT_{in}=0.64\pm0.09$keV, $kT_{o}=0.45^{+0.06}_{-0.02}$keV,
$kT_{e}=25\pm2$keV and $\tau=3.8\pm0.3$.\\
{\em {4U 0614+091}}: 
 In the \sax\/ data a feature at 0.7 keV, 
well reproduced by a Gaussian component, has been previously reported by Piraino et al. 1999.
Juett et al. (2001) 
attributed this feature in the \asca\/ spectra to an excess photoelectric absorption
due to a non solar abundance of Ne.  
Grating spectra with \chandra\/ and \xmm\/ confirm the unusual Ne/O ratio (Paerels et al. 2001).
Motivated by the above, we fitted again data with a new absorbed black body and Comptonization model,
allowing the relative abundances of O and Ne to vary (using the {\em {vphabs}} model in {\it {XSPEC}}).
We got for this fit a $\chi^{2}/$d.o.f.\ of 221/155, statistically similar to one with Gaussian component ($\chi^{2}/$d.o.f.=216/154)
and Ne and O abundance ratio relative to solar are $\sim$2.6 and $\sim$0.3, respectively, in agreement
with previously values reported by Juett al al. 2001.

\section{Discussion}
We have performed a systematic study of the average properties of the persistent
burster UCXBs. 
The IBIS long monitoring results indicate that UCXB
sources spend most of the time in the canonical low/hard state,
with unabsorbed luminosities lower than $\lesssim~8\times10^{36}~erg~s^{-1}$
but for the source 4U~1820-30 which was always detected in a 
soft/high state (see discussion below). \\
%sarebbe carino mettere le medie pesate ....ma !!!!\\
Figure \ref{model} clearly shows the majority of the sources are in the hard state
with constant emission 
between 10 and 70 keV. Generally, the hard component, 
due to the Comptonization of soft photons coming from
the accretion disk and/or the neutron star surface, dominates the emission.

\begin{figure*}[h]
\begin{center}
{\includegraphics[height=9cm, width=10cm, angle=-90]{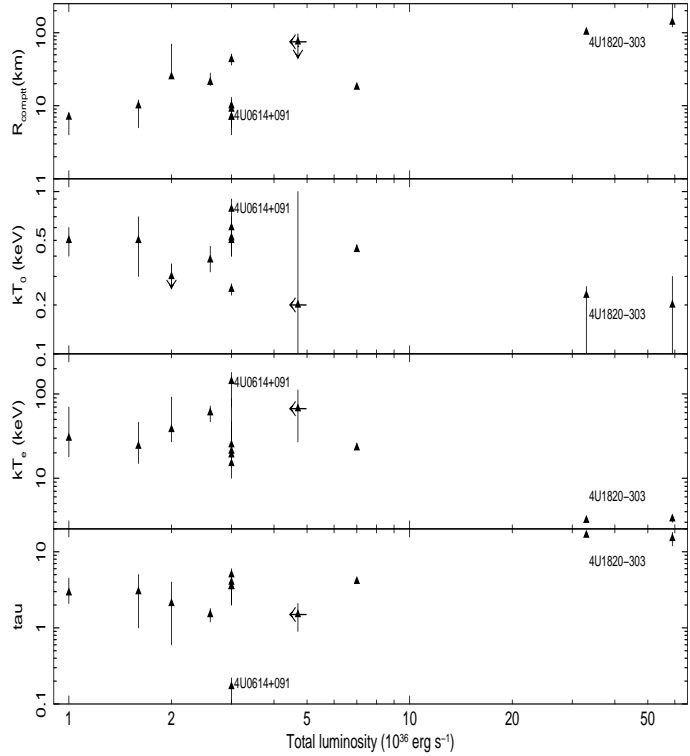}}
\caption{The radius of the emission area of the seed photons, the seed photons temperatures, 
the electron temperature and the optical depth of the Comptonizing plasma versus the total luminosity 
in the 0.1-100 keV energy range, in units of $10^{36}~erg~s^{-1}$}
\label{par}
\end{center}
\end{figure*}

The two sources 4U~1820-30 and 4U0614+091 have shown a different energy spectrum shape.
The first burster is observed only in a typical high/soft state with 
an accretion rate of $(0.6-1.0)\times10^{-8}\msun/yr$, a
low electron temperature ($\sim$~3~keV) and a very high optical depth ($\sim$15-16).
This behavior differs from other UCXB burster. We note that this
source could be a triple system consisting of a white dwarf 
accreting onto a neutron star
and a third body with mass weakly constrained ($0.004~\msun<M<0.5~\msun$),
with two timing modulations, the orbital period of 685s and super orbital variability
of $\sim$ 170 d (Zdziarski et al. 2007a, 2007b).
The second burster 4U~0614+091 shows a stronger disk emission
than other UCXB bursters and a very high electrons temperature ($\sim$~140~keV). 
This very high electrons temperature could be due to 
inverse Compton of seed photons with a not Maxwellian electrons distribution,
which generate
a power law spectrum without high energy cutoff as 
previously reported by Piraino et al. (1999).

The soft component of our sample is successfully fitted with a multi disk black body model,
providing reasonable physical parameters of the disk.
The temperatures at the inner disk radius is 0.1-0.6 keV 
(in one only case is $\sim 1.2)$, lower than 1~keV as previously reported 
by Sidoli et al. 2001 for a sample of 3 UCXBs.
Moreover, these authors show that three ultra compact binaries display 
seed photons temperatures $kT_0$ consisting with the temperature of
the inner regions of the accretion disk.
Although a clear correlation between the temperature
of the seed photons and the inner disk temperature  
is not observed for our sample, we note that values of both temperatures 
are in the range 0.1-0.8 keV, but for the candidate triple system 4U~1820-303. 
This behavior indicated that the inner disk could be 
identify as the region for the seed photons of the Comptonization emission. 

The thermal Comptonization component well fit the UCXB sample spectra in the 
high energy region. The electrons temperatures $kT_e$ are 
typically $\gtrsim~20$ keV, the optical depth $\tau~\lesssim~4-5$ and 
the seed photons temperatures $kT_0~\lesssim~0.8$ keV (but for 4U~1820-303 which is in a high/soft state). The relations between the {\scriptsize{COMPTT}} parameters obtained 
from the spectral analysis and the total luminosities are shown in figure \ref{par}.
As previously reported by Sidoli et al. 2001, the radius of the emission area of the
 seed photons seems to correlate with the total luminosity,
suggesting that the seed photons originate from the boundary layer 
rather than from the accretion disk (Popham and Sunyaev 2001).

Finally we note that 
the intermediate long burst (decay times ranging between ten and a few tens of minutes)
or superbursts (decay times more than an hour) are very rare events. 
Indeed
the estimated fraction of intermediate long burst
or superbursts among the  whole type I burst population is only 0.3-0.4\%.
In our studied sample 64\% of the UCXBs show superbursts or intermediate long burst
(as reported in table 1). This suggests that such timing
behavior could be a common feature to the UCXBs.
On the other hand, there are only 13 objects showing long or intermediate bursts
(Lewin and van der Klis, 2006) and 
they are UCXBs or binary systems with unknown orbital period. 
\begin{acknowledgements}
We acknowledge the ASI financial/programmatic support via contracts ASI-IR
I/008/07/0 and I/088/06/0.
A special thank to M. Federici for supervising
the \integral\/ data archive.
\end{acknowledgements}


\begin{thebibliography}{}
\bibitem[]{barr}Barret et al. 2003,  A\&A, 400, 643
\bibitem[]{bass} Bassa et al., 2006, A\&A, 446, L17
%\bibitem[]{baz} Bazzano et al. 1997, IAUC, 6668, 2
%\bibitem[]{bi98} Bildsten, L. 1998, in “The Many Faces of Neutron Stars” ed. R. Buccheri, J.
%van Paradijs, \& M. A. Alpar (Dordrecht: Kluwer), p. 419 
\bibitem[]{bi07}
Bird A. J., et al., 2007, \apj~S, 170, 175
\bibitem[]{blo} Bloser et al. 2000, ApJ, 542, 1000
\bibitem[]{boll} Boller T., Haberl F., Voges W., Piro L., Heise J., 1997, IAUC, 6546, 1
\bibitem[]{brad} Brandt et al. 1992, A\&A, 262, 115
%\bibitem[]{br06} Brandt, 2006, ATel 778
%\bibitem[2000]{ca:00}
%Campana S., 2000, ApJ 534, L79
\bibitem[]{chab} Chaboyer et al., 2000, AJ, 120, 3102
\bibitem[]{chev} Chenevez J.et al., 2007, A\&A, 469, 27
\bibitem[]{cocc} Cocchi et al., 2000, A\&A, 357, 527
\bibitem[]{coc1} Cocchi et al., 2001, MnSAI, 72, 757
\bibitem[]{corn} Cornelisse et al., 2002, A\&A, 392, 885
\bibitem[]{disk}
Dickey \& Lockman, 1990, Ann. Rew. Astron. Astrophys., 28, 215
\bibitem[1994]{eb:94}
Ebisawa, K., et al. 1994, PASJ, 46, 375
\bibitem[]{fala} Falanga et al. 2007, astroph 0711.0328, submitted A\&A
\bibitem[]{fiocchi} Fiocchi M. et al. 2008, A\&A, 477, 239
\bibitem[Fiore et al. 1999]{fio99}
Fiore, F., Guainazzi, M., \& Grandi, P. 1999,
Cookbook for BeppoSAX NFI Spectral Analysis
(www.asdc.asi.it/bepposax/software/cookbook)
%\bibitem[Fuj]{fuj} Fujimoto M. Y., Hanawa T., Miyaji, S., 1981, ApJ, 247, 267
%\bibitem[Fus]{fus} Fushiki I. and Lamb D. Q., 1987, ApJ, 323, 55
%\bibitem[]{gal} Galloway K., Muno M. P., Hartman J. M., Savov P., Psaltis D. , Chakrabarty D. astro-ph/0608259
%\bibitem[1979]{gl:79}
%Ghosh P., Lamb F.K., 1979, ApJ 234, 296
%\bibitem[1992]{gl:92}
%Ghosh P., Lamb F.K., 1992, in  van  den Heuvel E.P.J.,  Rappaport S.A. (eds.)
%``X--ray Binaries and Recycled Pulsars'',  (Dordrecht: Kluwer), p.~487
\bibitem{glad} Gladstone, J., Done, C., Gierliński, M., 2007, MNRAS, 378, 13
%\bibitem{goldw} Goldwurm, A., et al., 2004, ESASP, 552, 237
\bibitem{harri} Harris et al., 1996, AJ, 112, 1487
\bibitem{hei01} Heinke, C.O., Edmonds, P.D., \& Grindlay, J.E. 2001, ApJ, 562, 363
%\bibitem{hof80} Hoffman et al. 1980, ApJ, 240, 27
\bibitem{homer96} Homer, L., Charles, P.A., Naylor, T., et al. 1996,
        MNRAS, 282, L37
%\bibitem[]{ibe}Iben I. Jr., Tutukov A. V., 1991, ASPC, 20, 403
\bibitem[]{zan98} in 't Zand, J.J.M. et al., 1998, A\&A, 329, L37
\bibitem[]{zan99} in 't Zand, J.J.M., Heise, J., Bazzano, et al. 1999, IAUC 7243
\bibitem[]{zan02} in 't Zand, et al. 2002, A\&A, 389, L43
\bibitem[]{zan05} in 't Zand, J.J.M. et al., 2005, A\&A, 440, 287
\bibitem[]{zan07}
in 't Zand, J.J.M., Jonker, P. G., \& Markwardt C. B., 2007, A\&A, 465, 953
\bibitem[]{jonke} Jonker et al., 2004, 354, 355
%\bibitem[]{joss77} Joss, P. C., 1977, Nature, 270, 310
%\bibitem{joss78}Joss P. C., Avni Y., Rappaport S., 1978, ApJ, 221, 645
\bibitem{juett01}Juett A. M., Psaltis D., Chakrabarty D., 2001, ApJ, 560, 59
\bibitem[]{lew} Lewin and van der Klis M. 2006. In Compact Stellar X-ray Sources, ed. WHG Lewin, M van der Klis, Cambridge University Press
%\bibitem[]{loc} Lochner J.C., Roussel-Dupre D., 1994, ApJ, 435, 840 
\bibitem{kong} Kong A.K.H., 2006, ATel 875.
\bibitem[]{kuu}
Kuulkers et al. 2003, A\&A, 399, 663
%\bibitem[]{kuu05} kuulkers, 2005, ATel, 483
%\bibitem[]{mac} Machin et al., 1990, MNRAS, 247, 205
\bibitem[1986]{m:86}
Makishima K., Maejima Y., Mitsuda K., et al., 1986, ApJ 285, 712
%\bibitem[1984]{m:84}
%Mitsuda K., Inoue H., Koyama K., et al., 1984, PASJ 36, 741
\bibitem[]{mol} Molkov S., et al., 2005, A\&A, 434, 1069
%\bibitem[]{mor} Moretti et al., 2004, SPIE, 5165, 232
%\bibitem[]{nar} Narita et al., 2003, AJ, 593, 1007
\bibitem{nel03} Nelemans G., Jonker P. G., Marsh T., van der Klis M., 2004, MNRAS, 348, 7.
%\bibitem{nel06}Nelemans G. and Jonker P. G., 2006, astro-ph 5722.
\bibitem{nel06b}Nelemans G., Jonker P. G., Steeghs D., 2006, MNRAS, 370, 255
\bibitem{ne86}Nelson L. A., Rappaport S. A., Joss P. C., 1986, ApJ, 311, 226
\bibitem{obr05} O'Brien, K., presentation at symposium 'A Life With
        Stars', Amsterdam, 2005
\bibitem[]{pae}Paerels et al. 2001, ApJ, 546, 338
\bibitem[pal]{pal}Paltinieri et al., 2001, AJ, 121, 3114
\bibitem[]{paR01} Parmar A. N., Oosterbroek T., Sidoli L. et al. 
2001, A\&A, 380, 490
\bibitem[]{pir99}
Piraino S., Santangelo A., Ford E. C., Kaaret P., 1999, A\&A, 349, 77.
\bibitem[]{piro}
Piro et al. 1997, IAUC, 6538, 2
%\bibitem[1986]{p:86}
%Priedhorsky W., 1986, ApJ 306, L97
%\bibitem[{Podsiadlowski} et~al.(2002)]{prp02}
%Podsiadlowski P., Rappaport S., and Pfahl E.D., 2002, \apj, 565, 1107
\bibitem[]{pop} Popham R. \& Sunyaev R., 2001, \apj, 547, 355
%\bibitem[]{sav} Savonije G. J., de Kool, M., van den Heuvel E. P. J., 1986, A\&A, 155, 51 
%\bibitem[]{sch} Schatz H., Bildsten L., Cumming A., Wiescher M., 1999, ApJ, 524, 1014
\bibitem[]{sid01} Sidoli L., Parmar A. N., Oosterbroek T. et al., 2001, 368, 451
\bibitem[]{sid} Sidoli et al. 2006, A\&A, 460,229
\bibitem[]{sha}
Shaposhnikov, N., Titarchuk, L.,  2004, ApJ, 606, 57
\bibitem[1995]{sh:95}
Shimura, T., \& Takahara, F. 1995, ApJ, 445, 780
%\bibitem[]{sid} Sidoli, L. et al., 2001, A\&A, 368, 451
%\bibitem[]{sma}Smale et al., 1988, MNRAS, 232, 647
\bibitem{ste86} Stella, L., Priedhorsky, W., \& White, N.E. 1987, ApJ, 312, L17
\bibitem{sroma} Strohmayer \& Brown, 2002, ApJ, 566, 1045
%\bibitem[]{taam} Taam R. E. and Picklum R. E., 1979, ApJ, 233, 327
%\bibitem[]{tara} Tarana et al. 2007, ApJ, 654, 494
\bibitem[1994]{ti:94}
Titarchuk L., 1994, ApJ 434 570
%\bibitem[]{}Tutukov A. V., Iungelson L. R., 1987, Ap\&SS, 130, 15
\bibitem[ube]{ube} Ubertini et al., 2003, A\&A, 411, 131 
\bibitem[]{vac} Vacca et al. 1986, 220, 339
\bibitem[{Van Paradijs} and {McClintock}(1994)]{vm94}
van Paradijs J. and McClintock J. E., 1994, \aap, 290, 133
\bibitem{wi}Winkler C. et al., 2003, A\&A, 411, 1
%\bibitem[]{yug} Yungelson, Nelemans and Van den Heuvel, 2001
%\bibitem{whi82} White, N.E., \& Swank, J.H. 1982, ApJ, 253, L61
\bibitem{zdz7a}
Zdziarski A. A., Gierliaski M., Wen L.
 2007a, MNRAS, 377, 1017
\bibitem{zdz7b}
Zdziarski A. A., Wen L., Gierliaski M.
2007b, MNRAS, 377, 1006
\end{thebibliography}
\end{document}